# Artificial intelligence and biological misuse: Differentiating risks of language models and biological design tools

Jonas B. Sandbrink[1*]

[1] *Nuffield Department of Medicine, University of Oxford, Oxford, United Kingdom*
*\* Correspondence: jonas.sandbrink@trinity.ox.ac.uk*

**Abstract**

As advancements in artificial intelligence (AI) propel progress in the life sciences, they may also enable the weaponisation and misuse of biological agents. This article differentiates two classes of AI tools that could pose such biosecurity risks: large language models (LLMs) and biological design tools (BDTs). LLMs, such as GPT-4 and its successors, might provide dual-use information and thus remove some barriers encountered by historical biological weapons efforts. As LLMs are turned into multi-modal lab assistants and autonomous science tools, this will increase their ability to support non-experts in performing laboratory work. Thus, LLMs may in particular lower barriers to biological misuse. In contrast, BDTs will expand the capabilities of sophisticated actors. Concretely, BDTs may enable the creation of pandemic pathogens substantially more devastating than anything seen to date and could enable forms of more predictable and targeted biological weapons. In combination, the convergence of LLMs and BDTs could raise the ceiling of harm from biological agents and could make them broadly accessible. A range of interventions would help to manage risks. Independent pre-release evaluations could help understand the capabilities of models and the effectiveness of safeguards. Options for differentiated access to such tools should be carefully weighed with the benefits of openly releasing systems. Lastly, essential for mitigating risks will be universal and enhanced screening of gene synthesis products.

**Introduction**

Artificial intelligence (AI) has the potential to catalyse enormous advances in the life sciences and medicine. However, as AI accelerates the life sciences, it may also enable harmful and malicious applications of associated capabilities. Urbina *et al.* have demonstrated how an AI-powered drug discovery tool could be used to generate blueprints for plausible novel toxic chemicals that could serve as chemical weapons [1]. Similarly, AI may also empower the weaponisation and misuse of biological agents — and because of their potentially transmissible nature, risks from biological agents may exceed that of chemical ones. As the tremendous potential benefits of artificial intelligence for the life sciences are already widely discussed, this article focuses explicitly on creating a taxonomy of different potential biosecurity risks of AI systems. Many of the discussed risks are still speculative, but the identification of such potential risks will allow researchers and security experts to collect evidence on their existence and severity.

This article specifically differentiates two forms of AI which, in different ways, could exacerbate biosecurity risks: large language models (LLMs) and biological design tools (BDTs). I introduce the term biological design tools to describe systems that are trained on biological data and can help design proteins or other biological agents. LLMs and BDTs feature significantly different properties and risk profiles.

Next to the direct ways in which these tools could enable the creation of biological weapons, AI systems may also increase biosecurity risks through indirect avenues. For instance, LLMs could exacerbate misinformation and disinformation challenges [2], which could negatively impact the response and attribution of a biological event. Furthermore, LLMs could serve as tools to radicalise and recruit or to coerce and manipulate scientists to share pathogen samples or acquire technical expertise for biological weapons development. These risks are less unique to biosecurity and are not the focus of this piece.

**Risks from large language models (LLMs)**

The first class of AI tools that might enable misuse of biology are large language models (LLMs) that have been trained on large amounts of text, including scientific documents and discussion forums. LLMs and related "AI assistants" can provide scientific information, access relevant online resources and tools, and instruct research. Examples include foundation models (e.g. GPT-4/ChatGPT), language models optimised for assisting scientific work (e.g. BioGPT) [3], and LLM-based applications for interfacing with other scientific tools and laboratory robots [4,5]. While foundation models are products of large and expensive training runs and are currently developed by a small number of companies, these systems can be optimisied and turned into applications for particular domains by more resource-constrained groups [4,5].

LLMs might impact the risks of biological misuse in several ways. A key theme is that LLMs could increase the accessibility to existing knowledge and capabilities, and thus may lower the barriers to biological misuse (see Figure 1b).

|  | without LLMs | with LLMs |
|---|---|---|
| **without BDTs** | *Status quo* — Most individuals are not able to access biological agents, and only a small number of actors are capable of causing large-scale harm. | Existing capabilities accessible to a larger number of individuals |
| **with BDTs** | Increased ceiling of capabilities | Increased the ceiling of capabilities accessible to a larger number of individuals |

**Figure 1: Potential effects of LLMs and BDTs on capabilities for biological misuse**
Illustrative schematic of how artificial intelligence tools could impact capabilities across the spectrum of actors with the potential to misuse biology, if no safeguards are deployed. LLMs = large language models; BDTs = biological design tools.

*1. Teaching about dual-use topics*
First, LLMs may enable efficient learning about "dual-use" knowledge which can be used for informing legitimate research but also for causing harm. While it remains an open question whether current LLMs are better than internet search engines at teaching about dual-use topics, LLMs feature properties that mean that current or future systems have this potential. In contrast to internet search engines, LLMs can synthesise knowledge across many different sources, make complex information accessible and tailored to non-experts, and can proactively point out variables that the user did not know to inquire about. If biological weapons-enabling information is presented in this way, this could enable smaller biological weapons efforts to overcome key bottlenecks. For instance, one hypothesised factor for the failed bioweapons efforts of the Japanese doomsday cult Aum Shinrikyo is that its lead scientist Seichii Endo, a PhD virologist, failed to appreciate the difference between the bacterium *Clostridium botulinum* and the deadly botulinum toxin it produces [6]. At time of testing in June 2023, ChatGPT readily outlines the importance of "harvesting and separation" of toxin-containing supernatant from cells and further steps for concentration, purification, and formulation. While there are also other reasons why Aum failed to produce biological weapons [7], this example illustrates how a chatbot could provide helpful information to potential bioterrorists. Similarly, LLMs might have helped Al-Qaeda's lead scientist Rauf Ahmed, a microbiologist specialising in food production, to learn about anthrax and other promising bioweapons agents, or they could have instructed Iraq's bioweapons researchers on how to successfully turn its liquid anthrax into a more dangerous powdered form [6,8]. Current LLMs still produce at times "hallucinations" in the form of false or made-up information, which may frustrate or mislead potential malicious actors; however, such hallucinations will likely disappear as model developers develop increasingly optimised systems.

*2. Identifying specific avenues to biological misuse*
Second, LLMs may help with the ideation and planning of how to attain, modify, and disseminate biological agents. Already, LLMs are able to identify how existing supply chains can be exploited to illicitly acquire biological agents. In a recent one-hour exercise, LLMs enabled non-scientist students to identify four potential pandemic pathogens and how they can be synthesised, which companies supply synthetic DNA without screening customers and orders, and the potential to engage contract service providers for relevant laboratory work [9]. However, this experiment was not controlled with a group attempting to find the same information on the internet. More carefully controlled studies may shed more light on how much uplift current LLMs are providing compared to the *status quo* [10]. However, in the longer term, LLMs could also generate ideas for how to design biological agents tailored for a specific goal, such as what molecular targets would be best suited to produce a particular pathology.

*3. Step-by-step instructions and trouble-shooting experiments*
Additionally, LLMs could become very effective laboratory assistants which might be able to provide step-by-step instructions for experiments and guidance for troubleshooting experiments. Particularly important for this could be the advent of multi-modal models, which will not only be able to give personalised written instructions but also will be able to produce images and video. Such AI lab assistants will have many beneficial applications for helping less experienced researchers and replicating experimental methods from publications. However, these AI lab assistants might also support laboratory work for malicious purposes. For instance, a key reason for Aum Shinrikyo's failure to weaponise anthrax was that Seiichi Endo did not succeed at turning a benign vaccine strain of the bacterium into its pathogenic form, despite access to relevant protocols for plasmid insertion. Endo might have succeeded with an AI lab assistant to provide tailored instructions and help with troubleshooting. One crucial open question is the barrier formed by "tacit knowledge", knowledge that is generally not well-documented or cannot easily be put into words [11]. Different categories of tacit knowledge have been defined [12], which may be impacted by AI lab assistants to different degrees. "Somatic" tacit knowledge, such as the muscle memory of how to use a pipette, may remain largely untouched by AI laboratory assistants. In contrast, "weak" tacit knowledge, such as tweaks to lab protocols that are not well-documented but can be put into words, and "communal" tacit knowledge, which emerges from the confluence of many knowledge in different areas of expertise, have the potential to be lowered by AI lab assistants that have been optimised to provide tailored laboratory instructions and can draw on knowledge across many different disciplines. Lastly, if AI lab assistants create the perception that performing a laboratory feat is more achievable, more groups and individuals might try their hand — which increases the risk that one of them actually succeeds.

*4. Autonomous science capability*
In the longer term, as LLMs and related AI tools improve their ability to do scientific work with minimal human input, this could remove relevant barriers to biological weapons. Firstly, LLMs can instruct laboratory robots based on natural language commands, which will make them easier to use [13]. As lab robots become more advanced, this may reduce the need for human-conducted lab work and thus remove "somatic" tacit knowledge barriers. Secondly, LLMs can serve as the basis for autonomous science agents, which break tasks into manageable pieces, interface with relevant specialised computational tools, and instruct laboratory robots [5]. Challenges relating to coordinating large teams under secrecy limited the Soviet and Iraq bioweapons programs and likely has also served as a barrier for terrorist groups [6]. If autonomous science capabilities enable individuals and small groups to achieve large-scale scientific work, this will likely empower covert bioweapons programs.

**Risks from biological design tools (BDTs)**

The second class of AI tools that might pose a risk of misuse are biological design tools (BDTs). These BDTs are trained on biological data and can help design new proteins or other biological agents. Next to BDTs, there are also a range of other AI-enabled biological tools [14]. Examples of BDTs include RFDiffusion, as well as protein language models like ProGen2 and Ankh [15–17]. . In the future, LLMs finetuned on biological data could become the most powerful BDTs. Currently, BDTs are frequently open-sourced, regardless of whether they are developed by academia (RFDiffusion) or industry (ProGen2, Ankh). Next to tools for protein or organism design, there are also other machine learning tools with related dual-use implications, such as tools that shed light on host-pathogen interactions through predicting properties like immune evasion [18] or through advancing functional understanding of the human genome [19]. At the moment, biological design tools are still limited to creating proteins with relatively simple, single functions. However, eventually, relevant tools likely will be able to create proteins, enzymes, and potentially even whole organisms optimised across different functions.

There are three key ways in which BDTs might impact risks of biological misuse. In contrast to LLMs which may in particular increase the accessibility of biological weapons, BDTs may increase the ceiling of capabilities and thus the ceiling of harm posed by biological weapons (see Figure 1c). However, it is worth noting that BDTs may in the future also lower barriers to misuse, in particular if they end up producing zero-shot designs that do not require laboratory validation.

*1. Sophisticated groups and increased worst-case scenario risks*
First, as biological design tools advance biological design, this will likely increase the ceiling of harm that biological misuse could cause. It has been hypothesised that for evolutionary reasons naturally emerging pathogens feature a trade-off between transmissibility and virulence [20]. BDTs might enable overcoming this trade-off and allow the creation of pathogens optimised across both of these properties. Such pathogens might be released accidentally or deliberately, including by groups like Aum Shinrikyo. Bioterrorism with such designed pathogens is a low-probability scenario, because very few people have relevant motivations and - even with AI tools - designing an optimised pathogen will require significant skills, time, and resources. However, these barriers to using BDTs will decrease with advances in large language models and other AI lab assistants. Thus, humanity might face the threat of pathogens substantially worse than anything nature might create, including pathogens capable of posing an existential threat.

*2. State actors and new capabilities*
Second, biological design tools may be a key contributor to raising biological engineering capabilities in a way that makes biological weapons more attractive for state actors. The United States never included bioweapons developed during the 1960s in its war plans due to their short shelf life and the risk of harming friendly troops [6]. Iraq never deployed its bioweapons, likely because of a lack of certainty around its effectiveness and fear of retaliatory measures. If AI tools push the ceiling of biological design to make biological agents more predictable and targetable to specific geographic areas or populations, this could increase the attractiveness of biological weapons.

*3. Circumventing sequence-based biosecurity measures*
In the near-term, biological design tools will challenge existing measures to control access to dangerous agents. Examples include the taxonomy-based Australia Group List for export controls and the genetic sequence-based screening of synthetic DNA products. BDTs will likely make it easier to design potentially harmful agents that do not resemble the function or sequence of any known toxin or pathogen. Additionally, current or near-future inverse folding tools like ProteinMPNN [21] may already allow the "recoding" the structure of a known toxin in a substantially different genetic sequence. It is yet unclear how difficult it is to retain the original function of structures recoded in this way; likely results will be difficult to predict as suggested synonymous codon changes in poliovirus leading to loss of viral function [22]. As AI enhances biodesign capabilities, taxonomy or sequence similarity-based controls will not be sufficient to prevent illicit access to harmful biological agents in an age of AI-powered biological design.

**Takeaways for risk mitigation**

This characterisation of potential biosecurity risks of LLMs and BDTs can help investigations of such risks and catalyse the pre-emptive exploration of guardrails. To date, the described risks are still largely speculative. Robust evidence needs to be collected to identify whether LLMs are already lowering barriers to biological weapons and how these risks will evolve in the future. Particularly important for collecting such evidence will be studies carefully designed to evaluate how much uplift AI systems can provide compared to using the internet. Additionally, model developers and policymakers should pre-emptively consider what guardrails could be appropriate to mitigate future risks. This in particular might be important as AI capabilities may advance very fast and unpredictably [23]. One crucial area to follow is how LLMs interact with BDTs to make advanced biological design capabilities more accessible (see Figure 1d). Possible mechanisms include LLMs providing natural language interfaces to using BDTs, AI lab assistants helping to turn biological designs into physical agents, or eventually LLMs becoming more powerful at biological design than specialised tools.

*Pre-release model evaluations can characterise risks and effectiveness of safeguards*
Pre- and post-release model evaluations could be critical for assessing and mitigating risks of cutting-edge AI systems. [24]. OpenAI performed a prototype version of such pre-release model evaluations before the release of GPT-4 [25]. Ideally, pre-release model evaluations would involve an external and independent audit of foundation models. This could involve both expert red-teaming and more structured tests to evaluate model risks and safeguards, including relating to the ability to help with planning or execution of a biological attack. This would incentivise developers to remove harmful model behaviour throughout training and deployment. Pre-release model evaluations needs to involve various scaffolding, tooling, and finetuning of models to evaluate how system modifications could result in new capabilities.

*Careful consideration of trade-offs between unrestricted access and security*
The misuse potential of AI systems raises the difficult question of how to trade off open science and security [26]. The open release of a model, including publication of its weights, generally provides the least friction for further advances and in some instances can help improve the understanding of risks. At the same time, if model weights are published, any safeguards implemented by model developers can be removed and harmful versions of the model circulated [27]. Model developers and policy makers need to collect the evidence and develop the tools to understand when the benefits of openly releasing a model is outweighed by its risks. Where models are not openly released, a similar question needs to be addressed when deciding who to give access to a model. On the one hand, giving particular actors exclusive access to certain AI capabilities could exacerbate and harden social inequality. At the same time, indiscriminate access to AI systems that are capable of instructing non-experts on how to create a pandemic-capable virus could pose significant risks. Discussions involving a diverse range of stakeholders are needed on whether increasingly powerful lab assistants or BDTs should involve some kind of user authentication.

*Mandatory gene synthesis screening*
A range of interventions may be useful to mitigate risks from LLMs and BDTs [28]. One particularly effective way to mitigate increased risks from LLMs and BDTs might be to strengthen biosecurity measures at the boundary from the digital to the physical. Access to synthetic DNA is critical for translating any biological design into a physical agent. Industry leaders are already voluntarily screening gene synthesis orders and are calling for a regulatory baseline [29]. Such a mandatory baseline for the screening of gene synthesis orders and other synthetic biology services would be a very effective measure to prevent illicit access to biological agents. The 2023 US Executive Order on AI takes a first step towards creating such a baseline, by making the use of screened DNA mandatory for federally funded research [30]. At the same time, screening tools need to be improved in step with advances in biological design. For example, it will be important for future synthesis screening tools to predict protein structures and compare them to structures of known harmful agents. To this end, AI developers, biosecurity experts, and companies providing synthesis products could collaborate to develop appropriate screening tools.

## Conclusion

It is yet uncertain how and to what extent advances in artificial intelligence will exacerbate biosecurity risks. However, given the potential for fast advances in AI capabilities, it is prudent to adopt measures to assess and mitigate risks. If risks from AI can be effectively mitigated, this sets the groundwork for enabling AI to realise its very positive implications for the life sciences and human health.

## Acknowledgements

Markus Anderljung, Anemone Franz, Nicole Wheeler, and others for comments on the manuscript and helpful discussions. This piece only represents the opinion of the author, and not that of any of the organisations that they are working with. The author's doctoral research is funded by Open Philanthropy.

development-and-use-of-artificial-intelligence/